# Biological Insights from Integrative Modeling of Intrinsically Disordered Protein Systems

*Zi Hao Liu[1,2], Maria Tsanai[3,4], Oufan Zhang[3,4], Teresa Head-Gordon[3,4,5,6], Julie Forman-Kay[1,2,*]*

[1]Molecular Medicine Program, Hospital for Sick Children, Toronto, Ontario M5G 0A4, Canada
[2]Department of Biochemistry, University of Toronto, Toronto, Ontario M5S 1A8, Canada
[3]Pitzer Center for Theoretical Chemistry, University of California, Berkeley, California 94720, United States of America
[4]Department of Chemistry, University of California, Berkeley, California 94720-1460 United States of America
[5]Department of Chemical and Biomolecular Engineering, University of California, Berkeley, California 94720-1462, United States of America
[6]Department of Bioengineering, University of California, Berkeley, California 94720-1762, United States of America

*Author to whom correspondence should be addressed: forman@sickkids.ca

**Abstract**
Intrinsically disordered proteins and regions are increasingly appreciated for their abundance in the proteome and the many functional roles they play in the cell. In this short review, we describe a variety of approaches used to obtain biological insight from the structural ensembles of disordered proteins, regions, and complexes and the integrative biology challenges that arise from combining diverse experiments and computational models. Importantly, we highlight findings regarding structural and dynamic characterization of disordered regions involved in binding and phase separation, as well as drug targeting of disordered regions, using a broad framework of integrative modeling approaches.
**Keywords:** intrinsically disordered proteins, dynamic proteins, molecular dynamics, biomolecular condensates

**Introduction**

The abundance of protein sequences that do not adopt a stable 3-dimensional fold cannot be ignored given their high percentages in the proteome of various organisms. Disordered protein regions of at least 30 consecutive amino acids exist in ~60% of the human proteome [1], as predicted by SPOTDisorder [2]. Other work reports that 52-67% of eukaryotic proteins have IDRs of at least 40 consecutive amino acids long [3]. As such, intrinsically disordered proteins and regions (IDPs and IDRs) define an area of structural biology where there is still much to be learned about function from the structural ensembles that they adopt. Although there have been numerous recent advances in protein structure prediction tools such as AlphaFold2 [4], the primary focus has been on proteins that adopt a well-defined tertiary fold [5–8], leading to tools that are not appropriate for IDRs/IDPs [9]. Thus, new computational methods and experimental data are required to unravel the conformational ensembles of IDPs/IDRs.

Increasingly, robust pipelines for modeling single-chain IDPs are being reported [10–13], but, while ~5% of human proteins are predicted to be fully disordered, most disorder is within IDRs in the context of proteins having folded domains [1]. Furthermore, a large fraction of IDRs function via dynamic interactions with folded domains and other IDRs in either discrete dynamic complexes or within condensed states [14–17]. At present, there are far fewer approaches for modeling IDRs within the context of folded regions and dynamic complexes with other disordered or folded proteins, and in this review we illustrate the complexity of the full range of disordered systems beyond the IDP monomer.

**Integrative Modeling of Disordered Protein Systems**

Generating structural ensembles of isolated disordered chains, as summarized in Figure 1, involves using experimental data to compare with or to filter (subset or reweight) an initial diverse pool of possible structural conformers.

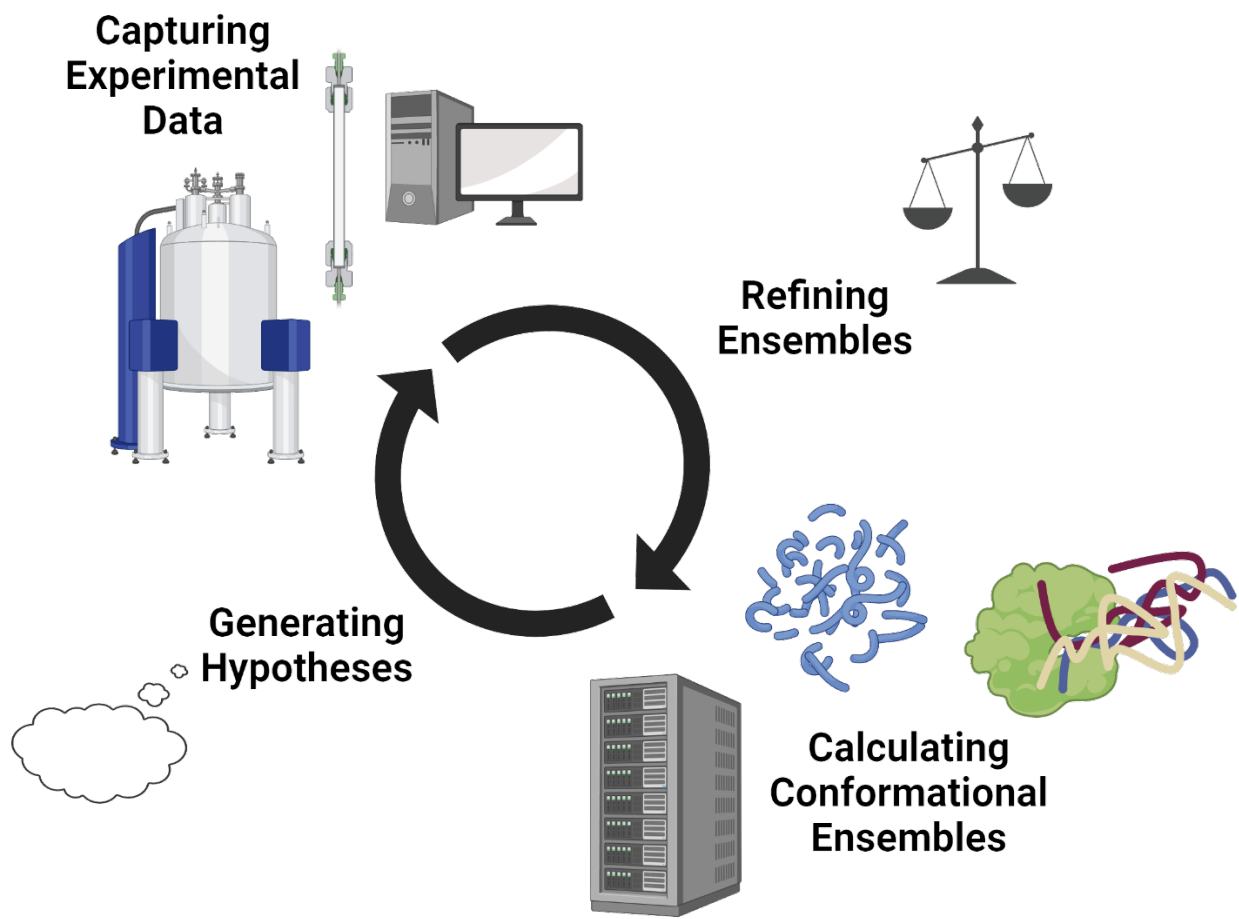

**Figure 1. Integrative modeling process of intrinsically disordered protein systems.** A schematic representation of experimental data obtained to funnel into a computational pipeline for ensemble generation, with both structural ensembles and experimental data assisting with hypothesis generation and illuminating the dynamics of the protein system.

Currently, there are three broad *in silico* methods to generate initial conformer pools: *i)* all-atom or coarse-grained molecular dynamics (MD) simulations [18–27], *ii)* generative machine learning (ML) models [28–32], and *iii)* statistical sampling of torsion angles and peptide fragments [11–13,33–36]. The conformers created by all of these methods can either be used *de novo*, can fit experiment by bias during conformer generation, or post-processing using sub-selection that agrees with experimental data [37–39]. While *de novo* MD simulations show reasonable agreement with the experiment in some cases, they also are known to have force field errors that make it challenging to fully match experiments [40,41]. This underscores the need to for an integrative biological approach in which experimental information from nuclear magnetic resonance (NMR), small angle X-ray scattering (SAXS) and other solution spectroscopies are used to bias simulated conformer generation [13,28].

The calculation of ensembles of disordered protein systems that closely reflect the available experimental data requires accurate back-calculators for experimental observables. Current back-calculators include those for NMR chemical shifts [42–45], paramagnetic relaxation enhancements (PREs) [46], and residual dipolar couplings (RDCs) [47]; SAXS [48,49]; single molecule fluorescence (SMF) [50,51] and electron paramagnetic resonance (EPR) [52]. Currently there is a need for more accurate back-calculators, which can be seen in the example of NMR chemical shift back-calculators which have root-mean-square errors that are much larger than differences expected for significantly different conformations [42,43]. This contrasts with the negligible typical experimental error for chemical shifts. One of the benefits of reweighting or sub-setting and ensemble scoring protocols is that it can account for the uncertainty of experimental and back-calculated values of observables, using Bayesian and other statistical methods [38,39]. Hence the current state of the art is the characterization of conformational ensembles of IDPs/IDRs by coupling structural modeling software with experimental data, yielding scientific insights into IDRs and their interactions. Various clustering tools can also identify ensemble sub-states that inform on local populations of conformations [53–56], and provide a means of gaining functional insight into the calculated ensembles.

**Integrative Modeling of Disordered Protein Regions**

*Regulation by post-translational modification of IDRs*

IDRs are the dominant sites of regulatory post-translational modifications (PTMs) on proteins, due to their accessibility [14]. Nucleosome core particles comprise the key unit of chromatin, and nucleosomal histones have significant IDRs which are the primary sites of PTMs, or epigenetic marks [57]. Lysine acetylation of the histone H4 tail (H4Kac) impacts chromatin architecture [58,59] by abolishing the 30 nm fiber formation *in vitro*, although the detailed mechanism has been debated [60]. MD simulations of the H4 tail were performed using the Amber03ws force field that strengthens protein-water interactions and is optimized for IDPs [25], together with lysine acetylation parameters derived from the PTM force field [26] providing AMBER parameters for 32 common PTMs. Secondary structure propensities [42] from NMR chemical shift data were used to verify simulations [61,62] of various multi-site acetylation (3Ac, 5Ac, unAc) states, supporting H4Kac leading to a shift in secondary structure propensity towards α-helix and β-sheet elements and a more compact ensemble due to increased intramolecular contacts between the basic patch and N-terminal region of H4 [63]. The authors postulate that compaction of the conformers could block H4KAc from forming higher order chromatin structure.

The intrinsically disordered 120-residue 4E-binding protein 2 (4E-BP2) inhibits eukaryotic mRNA cap-dependent translation by binding the eukaryotic initiation factor 4E (eIF4E) to block the formation of the translation initiation complex. Multi-site phosphorylation of 4E-BP2 stabilizes a ~40-residue β-sheet domain that sequesters the canonical binding helix and dramatically reduces binding to eIF4E to enable translation [64]. The phospho-sites in the C-terminal IDRs significantly stabilize the β-fold [65], but the mechanism is not clear. Tsangaris T.E. and co-authors [66] compared the non-phosphorylated 4E-BP2 IDP and the 5-fold phosphorylated 4E-BP2 with N- and C-terminal IDR tails around the β-domain. They used experimental data from single-molecule fluorescence resonance energy transfer (smFRET), SAXS, Cα/Cβ NMR chemical shifts, and PRE measurements, together with initial models derived from FastFloppyTail (FFT), which models chains using a three-residue-fragment based approach with a bias towards loop secondary structures [12], and the Bayesian/Maximum Entropy ensemble reweighting protocol that takes into consideration experimental error [38]. Back-calculation of FFT ensembles were done with AvTraj [51] for smFRET values, Pepsi-SAXS [49] for SAXS measurements, and ShiftX [45] for NMR chemical shifts. Analysis of the resulting ensembles by Cα-Cα Euclidean distance-based agglomerative clustering [67] demonstrated significant interactions of the β-domain with the N-terminal IDR, and showed that residues in the β-domain have fewer contacts with the C-terminal IDR than the non-phosphorylated 4E-BP2, providing models for stabilization that underlies regulation of translation.

For cell cycle progression in yeast, the Sic1 cyclin-dependent kinase is phosphorylated to enable binding and ubiquitination by SCF$^{Cdc4}$ with downstream degradation by the proteasome. The N-terminal (90 aa) IDR of Sic1 (here referred to as Sic1) and phosphorylated Sic1 (pSic1) was studied using NMR chemical shifts, PREs, SAXS, and smFRET [68]. Previously, only SAXS data were used to describe global dimensions of pSic1, but these authors calculated ensembles of Sic1 and pSic1, integrating a diverse collection of experimental observables, using ENSEMBLE [36], a Monte-Carlo algorithm to sub-set initial pools of conformers using various experimental datatypes. pSic1 was found to be more compact than Sic1, which enables electrostatic contributions from all phosphorylation sites to binding, yielding a sharp binding transition to Cdc4 as a function of number of phosphorylation sites [69,70]. The authors also found that ensembles jointly restrained by SAXS and NMR data were consistent with smFRET efficiencies that were not used in the refinement of the ensemble.

*IDRs involved in Biological Condensates*

IDRs are increasingly recognized for their ability to mediate phase separation contributing to formation of biomolecular condensates [71]. Galvanetto N. and Ivanović M.T. *et al.* explored dynamics of condensates formed by complex coacervation of the oppositely charged IDPs, human histone H1 and its nuclear chaperone, prothymosin-α (ProTα) [72], which function as chromatin condensation modulators. Experimental observations show that the ProTα-H1 dense phase is 1,000 times more concentrated than the dilute phase, leading to a bulk viscosity 300 times greater than water [72]; however, nanosecond fluorescence correlation spectroscopy (nsFCS) within the droplet reveals that the IDRs are dynamic on sub μs timescales. All-atom explicit-solvent MD simulations with the IDR-optimized Amber99SBws force field [24] and the TIP4P/2005s water model [27] were performed to characterize the dynamics of ProTα–H1 viscous coacervates [72]. Experimental nsFCS, smFRET, and NMR measurements were used to validate the simulations by comparing protein densities, diffusion coefficients, and FRET efficiencies, demonstrating reasonable agreement. Simulations of the dense phase showed rapid formation and breaking of individual contacts on the nanosecond time scale and revealed that the dense phase is formed by a network of rapidly rearranging and exchanging multivalent interactions between these oppositely charged proteins causing high macroscopic viscosity with fast molecular scale motions.

The C-terminal domain (CTD) of the RNA-binding protein TDP-43 is responsible for inducing phase-separation, although the direct contributions of specific residues are not clear. Mohanty P., Shenoy J., Rizuan A., *et al*. have used both atomistic and coarse-grained simulations together with critical saturation concentration ($c_{sat}$) and NMR measurements to identify the roles that aromatic and aliphatic residues in the hydrophobic conserved region (CR) play in driving phase separation [73]. The authors improved the accuracy of coarse-grained simulations by filtering residue-level contact profiles obtained from atomistic simulations using the Amber99SBws-STQ force field [24] with the TIP4P/2005s water model [27]. After ensuring that the simulations' secondary structure propensity are in agreement with NMR chemical shifts by using GROMACS [62] and DSSP [61], the trajectories were analyzed for residue specific contacts in and around the CR, yielding results that agreed with previous information on important contacts, and enabling validation of the roles of other residues probed by mutation.

The disordered, low-complexity ~100-residue C-terminal region of the human cytoplasmic activation proliferation-associated protein-1 (Caprin1) phase separates in the presence of various molecules, including salt and adenosine triphosphate (ATP). The full-length Caprin1 (709 residues) RNA-binding protein is found in a variety of cytoplasmic biomolecular condensates, facilitating its critical role in RNA processing [74]. Caprin1 dysfunction is linked to several pathologies, including nasopharyngeal carcinoma, intellectual disability, and autism spectrum disorder [75]. Lin Y. H. *et al.* combined theoretical and computational methods to investigate trends observed in NMR experiments [76,77] and their results reveal that interchain ion bridges enhance phase separation, while ATP acts both as a salt-like ion and an amphiphilic hydrotrope. The colocalization of (ATP-Mg)$^{2-}$ within condensates is driven by its high valency and ability to form electrostatic and π-related interactions. In a complementary study, using a computationally multiscale approach, Tsanai M. and Head-Gordon T. investigated the ATP-induced phase behavior of Caprin1 across three states: the initial mixed state, nanodroplet formation, and droplet dissolution [78]. Their findings that nanodroplets consist of stacked ATP clusters stabilized by sodium counterions and π-π interactions, which engage with the arginine-rich N-terminus of Caprin1, are consistent with residue-specific chemical shifts [78,79]. By calculating both the near-surface electrostatic potentials (NS-ESP) and the zeta potentials of Caprin1, a positive NS-ESP was observed during the initial mixed state, consistent with NMR data [76,77]. Interestingly, at the surface of the condensate a highly negative NS-ESP and significant zeta potentials were calculated outside the highly dense region of charge, which explain the remarkable stability of this phase separated droplet assembly [78,80]. At high ATP concentrations, weaker interactions drive the dissolution of droplets back into the mixed state exhibiting a much lower zeta potential.

The C-terminal IDR of the nucleoprotein (*N*) from the measles virus ($N_{TAIL}$) is essential for the formation of condensates that facilitate transcription and replication of the virus within infected cells [81]. By coupling NMR experimental data with MD simulations, Guseva S. and Schnapka V. *et al*. rationalized why the $N_{TAIL}$ displayed a separation of translational/rotational dynamics from dynamics from internal dynamics and showed how residue-specific contacts and their frequencies of contact modulate condensate viscosity [82]. $^{15}$N NMR relaxation experiments were used to measure the dynamics of the measles virus $N_{TAIL}$ in both the dilute and dense phase. An *in silico* model of the $N_{TAIL}$ was generated with ASTEROIDS [37], using PREs and RDCs to select conformations from an initial statistical pool, and then feeding these structures into all-atom MD simulations using the CHARMM36m forcefield [82]. Authors found that the dense phase slows down the translational dynamics of $N_{TAIL}$ compared to the dilute phase, however, the backbone conformational sampling, i.e., internal dynamics, is similar between the dilute and dense phases. The highly dynamic nature of the IDR within the dense phase is similar to the ProTα-H1 case mentioned previously [72]. However, unlike ProTα-H1, further all-atom MD simulations that support the NMR observables hint that the higher levels of crowding in the dense phase increase intermolecular contacts.

*IDRs in cellular regulation*

Statistical sampling methods have been used without experimental restraints for rapid generation of ensembles of large protein systems involving IDRs, for comparison with biological and biochemical data. One example is the eukaryotic condensin complex, a key regulator of chromatin condensation, chromosome assembly, and segregation during mitosis. Pastic A. *et al.* [83] investigated the mechanism of condensin complex targeting DNA, identifying a DNA-binding domain on the N-terminal IDR of the Smc4 core component of the complex responsible for chromatin interaction as well as influencing biomolecular condensate formation [84]. Since the full structural model of Smc4 has not been experimentally solved, the partial structure deposited in the RCSB PDB [85] and an AlphaFold 2 [86] model was used in conjunction with IDPConformerGenerator [13,34] to model ~15,000 all-atom conformers of the full-length Smc4 in the context of the condensin complex with DNA to predict the accessibility of the N-terminal DNA binding region [25]. The visual observation that the N-IDR of Smc4 could interact with the DNA supports the hypothesis that the N-terminal DNA binding domain is highly dynamic and is sterically unobstructed in the context of the condensin complex, facilitating its DNA-binding function.

Another modeling study illustrated how an IDR functions as a dynamic linker in regulation of apoptosis (Figure 2) [87]. This process depends on activation of caspase-9 (Casp9) protease through binding to the apoptosome (Apaf1) scaffold via its CARD binding domain. To estimate the effective concentration of the Casp9 protease domains (PDs), the IDR linker was modeled with 20,000 all-atom full-length Casp9 tetramers bound to the apoptosome using IDPConformerGenerator [13,34], a cryo-EM structure of the apoptosome [88] (containing the Apaf1 CARDs and Casp9 without coordinates for the IDR linker or PDs), and the AlphaFold 2 predicted structure of the Casp9 PD. Using UCSF ChimeraX [89], the effective concentration of the PD was estimated to be 560 µM, whereas the experimental effective concentration was found to be 470-560 µM [87], validating the calculated models, which, together with NMR and biochemical data, demonstrate that the PD functions at the end of a disordered tether, rather than being activated by interactions with the apoptosome.

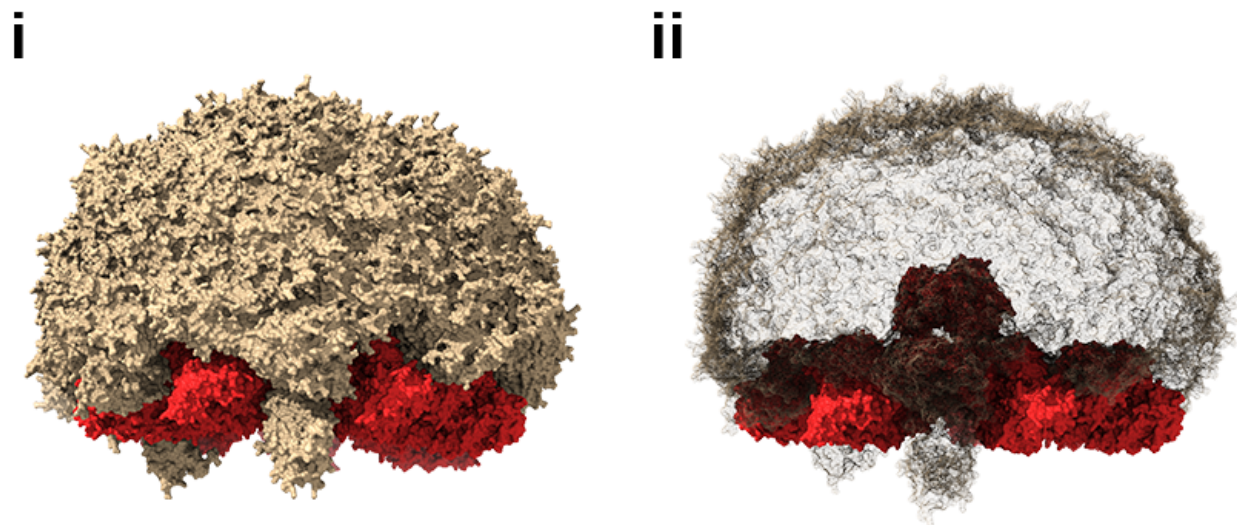

**Figure 2. Example of a large biological system with IDRs modeled with IDPConformerGenerator. i)** Structural models (N = 20,000) of caspase-9 protease domain [86] linked with IDRs (beige) generated by IDPConformerGenerator [13,34] attached to the caspase-9 CARD domains on top of the apoptosome highlighted in red (PDB: 5WVE). **ii)** Same figure as i) but with increased transparency of the caspase-9 IDR linker and protease domain highlighted in beige. Adapted from Sever A. I. M. and Alderson R. T., *et al*. [87].

*IDRs Ensembles for Drug Discovery*

IDR conformational ensembles have also been a target for drug discovery. Compounds that bound to ensembles of the N-terminal IDR of the oncoprotein p53 transactivation domain I (TAD1) were designed based on MD simulations, global clustering, and validation using NMR TOCSY experiments [90]. In another study, MD simulations using a force-field optimized for disordered proteins (a99SB-disp [19]) coupled with *in vitro* binding assays were used to design small molecule inhibitors that target the N-terminal IDR of the androgen receptor to treat castration-resistant prostate cancer [91].

**Perspectives**

Generating ensemble models of single-chain IDPs has become standard in recent years [11–13,33], especially with the refinement of MD and knowledge-based sampling techniques [18,19,22,92,93]. As the community moves to describe more complex systems, back-calculators will be needed that have higher accuracy [44] and are effective for multi-chain dynamic complexes and IDPs/IDRs with post-translational modifications (PTMs) [94]. Currently, a suite of back-calculators for analyzing conformational ensembles (chemical shift, $^3$J-coupling, PREs, nuclear Overhauser effect, smFRET, SAXS, $R_h$, and RDCs) can be found within the Structural Python Back-Calculator Interface for PDBs (SPyCi-PDB) for comparisons with different back-calculator methods [95]. Given the diversity of methods for calculating each experimental datatype, it would be worthwhile benchmarking tools to determine the best models to use for different applications.

With a variety of experimental techniques used to obtain data on IDRs/IDPs, there have been efforts, as seen in the Protein Ensemble Database (PED) [96], to maintain a unified database to host experimental data for disordered protein systems. However, these data are only deposited if connected to a calculated structural ensemble. While the Biological Magnetic Resonance Bank (BMRB) is a highly valuable database for NMR data, in practice not all described NMR experimental data types can be uploaded and there can be challenges of a unified format for the different datatypes. Additional relevant databases include the Small Angle Scattering Biological Data Bank (SASBDB) [97] as well as the Public Repository for Circular Dichroism Spectral Data (PCDDB) [98]. A database specific for DEER and EPR spectroscopy does not exist at the time of writing.

To assist with ML approaches, it would be ideal to simplify formatting to encourage more researchers to submit as much experimental data as possible, along with experimental conditions, to expand on the availability of accurate training data. Very recently, a number of ML models have been presented for generating IDP conformer ensembles, e.g., idpGAN [29], idpSAM [30], and IDPFold [31], with current application to reveal biological insights and further methodological development expected to benefit from increased data availability.

**Summary and Outlook**

Here we have highlighted examples of the increasing number of structural ensembles of IDRs and dynamic complexes, along with findings that inform physicochemical mechanisms underlying function derived from analysis of experimentally restrained conformer ensembles or unrefined ensembles that are compared with experiment. All the highlighted applications with biological insights should be of special interest in accordance with the journal's recommended reading guidelines; to avoid redundancy, we have instead annotated papers describing methodological approaches. With the measurement of more experimental data and the development of more accurate back-calculators and creative integrative modeling tools, much more detailed information will be discovered about IDRs and their dynamic complexes which regulate biology.

**Funding**

T.H.-G. and J.D.F.-K. acknowledge funding from the National Institutes of Health under Grant 2R01GM127627-05. J.D.F.-K. also acknowledges support from the Natural

Sciences and Engineering Research Council of Canada (RGPIN-2024-05725) and from the Canada Research Chairs Program. Z.H.-L. acknowledges funding from the Natural Sciences and Engineering Research Council of Canada (PGS D – 588933 – 2024).

**Declaration of competing interest**

The authors declare they have no competing financial interests that could have influenced the work reported in this article.

This publication describes CALVADOS 2, a coarse-grained model of IDPs, and its applicability to predict bulk conformational properties of IDPs including their propensities to undergo phase separation. The authors showcase the accuracy of the model for prediction of chain compaction, as well as phase separation propensities, across IDP cases with different lengths and charge patterning in different temperatures and salt conditions.

**22. Cao F, von Bülow S, Tesei G, Lindorff-Larsen K: A coarse-grained model for disordered and multi-domain proteins**. *Protein Science* 2024, **33**:e5172.

CALVADOS 3, an update to CALVADOS, is presented here, with the ability to included fixed positions for folded domains in the coarse-grain model to effectively model IDRs in the context of folded domains.

23. Ginell GM, Emenecker RJ, Lotthammer JM, Usher ET, Holehouse AS: **Direct prediction of intermolecular interactions driven by disordered regions**. 2024, doi:10.1101/2024.06.03.597104.

24. Lindorff-Larsen K, Piana S, Palmo K, Maragakis P, Klepeis JL, Dror RO, Shaw DE: **Improved side-chain torsion potentials for the Amber ff99SB protein force field**. *Proteins* 2010, **78**:1950.

25. Best RB, Zheng W, Mittal J: **Balanced Protein–Water Interactions Improve Properties of Disordered Proteins and Non-Specific Protein Association**. *J Chem Theory Comput* 2014, **10**:5113–5124.

26. Khoury GA, Thompson JP, Smadbeck J, Kieslich CA, Floudas CA: **Forcefield_PTM: Ab Initio Charge and AMBER Forcefield Parameters for Frequently Occurring Post-Translational Modifications**. *J Chem Theory Comput* 2013, **9**:5653–5674.

27. Abascal JLF, Vega C: **A general purpose model for the condensed phases of water: TIP4P/2005**. *J Chem Phys* 2005, **123**:234505.

*28. Zhang O, Haghighatlari M, Li J, Liu ZH, Namini A, Teixeira JMC, Forman-Kay JD, Head-Gordon T: **Learning to evolve structural ensembles of unfolded and disordered proteins using experimental solution data**. *The Journal of Chemical Physics* 2023, **158**:174113.

This publication reports a generative ML model, DynamICE, that uses input NMR experimental data to generate conformational ensembles that can fit experimental observables, based on an initial model that learned how to generate IDP structures from IDPConformerGenerator conformers.